\documentclass[twocolumn]{aastex631}

\usepackage{xspace}
\usepackage{multirow}

\newcommand{\halpha}{H$\alpha$\xspace}
\newcommand{\hii}{\ion{H}{2}\xspace}
\newcommand{\hone}{H{\sc i}\xspace}
\newcommand{\ration}{F1130W/F770W\xspace}
\newcommand{\ratioss}{F335M$_{\rm PAH}$/F770W\xspace}
\newcommand{\ratiose}{F335M$_{\rm PAH}$/F1130W\xspace}
\newcommand{\hagasratio}{$I_{\rm H{\alpha}}/\Sigma_{\rm H{\sc I}+H_2}$\xspace}
\newcommand{\rpah}{$R_{\rm PAH}$\xspace}
\newcommand{\mjysr}{MJy~sr$^{-1}$\xspace}

\shorttitle{PAH Band Ratios and ISM in PHANGS-JWST}
\shortauthors{Chastenet, Sutter, Sandstrom et al.}

\begin{document}

\title{PHANGS-JWST First Results: Measuring PAH Properties across the multiphase ISM}

\author[0000-0002-5235-5589]{J\'er\'emy Chastenet}
\affil{Sterrenkundig Observatorium, Ghent University, Krijgslaan 281-S9, 9000 Gent, Belgium}
\author[0000-0002-9183-8102]{Jessica Sutter}
\affiliation{Center for Astrophysics and Space Sciences, Department of Physics, University of California, San Diego\\9500 Gilman Drive, La Jolla, CA 92093, USA}
\author[0000-0002-4378-8534]{Karin Sandstrom}
\affiliation{Center for Astrophysics and Space Sciences, Department of Physics, University of California, San Diego\\9500 Gilman Drive, La Jolla, CA 92093, USA}

\author[0000-0002-2545-5752]{Francesco Belfiore}
\affiliation{INAF — Arcetri Astrophysical Observatory, Largo E. Fermi 5, I-50125, Florence, Italy}
\author[0000-0002-4755-118X]{Oleg V. Egorov}
\affiliation{Astronomisches Rechen-Institut, Zentrum f\"{u}r Astronomie der Universit\"{a}t Heidelberg, M\"{o}nchhofstra\ss e 12-14, 69120 Heidelberg, Germany}
\author[0000-0003-3917-6460]{Kirsten L. Larson}
\affiliation{AURA for the European Space Agency (ESA), Space Telescope Science Institute, 3700 San Martin Drive, Baltimore, MD 21218, USA}
\author[0000-0002-2545-1700]{Adam~K.~Leroy}
\affiliation{Department of Astronomy, The Ohio State University, 140 West 18th Avenue, Columbus, Ohio 43210, USA}
\affiliation{Center for Cosmology and Astroparticle Physics, 191 West Woodruff Avenue, Columbus, OH 43210, USA}
\author[0000-0001-9773-7479]{Daizhong Liu}
\affiliation{Max-Planck-Institut f\"ur Extraterrestrische Physik (MPE), Giessenbachstr. 1, D-85748 Garching, Germany}
\author[0000-0002-5204-2259]{Erik Rosolowsky}
\affiliation{Department of Physics, University of Alberta, Edmonton, Alberta, T6G 2E1, Canada}
\author[0000-0002-8528-7340]{David A. Thilker}
\affiliation{Department of Physics and Astronomy, The Johns Hopkins University, Baltimore, MD 21218, USA}
\author[0000-0002-7365-5791]{Elizabeth J. Watkins}
\affiliation{Astronomisches Rechen-Institut, Zentrum f\"{u}r Astronomie der Universit\"{a}t Heidelberg, M\"{o}nchhofstra\ss e 12-14, 69120 Heidelberg, Germany}
\author[0000-0002-0786-7307]{Thomas G. Williams}
\affiliation{Sub-department of Astrophysics, Department of Physics, University of Oxford, Keble Road, Oxford OX1 3RH, UK}
\affiliation{Max-Planck-Institut f\"{u}r Astronomie, K\"{o}nigstuhl 17, D-69117, Heidelberg, Germany}

\author[0000-0003-0410-4504]{Ashley.~T.~Barnes}
\affiliation{Argelander-Institut f\"{u}r Astronomie, Universit\"{a}t Bonn, Auf dem H\"{u}gel 71, 53121, Bonn, Germany}
\author[0000-0003-0166-9745]{F. Bigiel}
\affiliation{Argelander-Institut f\"ur Astronomie, Universit\"at Bonn, Auf dem H\"ugel 71, 53121 Bonn, Germany}
\author[0000-0003-0946-6176]{Médéric~Boquien}
\affiliation{Centro de Astronomía (CITEVA), Universidad de Antofagasta, Avenida Angamos 601, Antofagasta, Chile}
\author[0000-0002-5635-5180]{M\'elanie Chevance}
\affiliation{Institut f\"{u}r Theoretische Astrophysik, Zentrum f\"{u}r Astronomie der Universit\"{a}t Heidelberg,\\ Albert-Ueberle-Strasse 2, 69120 Heidelberg, Germany}
\affiliation{Cosmic Origins Of Life (COOL) Research DAO, coolresearch.io}
\author[0000-0002-5782-9093]{Daniel~A.~Dale}
\affiliation{Department of Physics and Astronomy, University of Wyoming, Laramie, WY 82071, USA}
\author[0000-0002-8804-0212]{J.~M.~Diederik~Kruijssen}
\affiliation{Cosmic Origins Of Life (COOL) Research DAO, coolresearch.io}
\author[0000-0002-6155-7166]{Eric Emsellem}
\affiliation{European Southern Observatory, Karl-Schwarzschild-Stra{\ss}e 2, 85748 Garching, Germany}
\affiliation{Univ Lyon, Univ Lyon1, ENS de Lyon, CNRS, Centre de Recherche Astrophysique de Lyon UMR5574, F-69230 Saint-Genis-Laval France}
\author[0000-0002-3247-5321]{Kathryn~Grasha}
\affiliation{Research School of Astronomy and Astrophysics, Australian National University, Canberra, ACT 2611, Australia}   
\affiliation{ARC Centre of Excellence for All Sky Astrophysics in 3 Dimensions (ASTRO 3D), Australia}   
\author[0000-0002-9768-0246]{Brent Groves}
\affiliation{International Centre for Radio Astronomy Research, University of Western Australia, 7 Fairway, Crawley, 6009 WA, Australia}
\author[0000-0002-8806-6308]{Hamid Hassani}
\affiliation{Department of Physics, University of Alberta, Edmonton, Alberta, T6G 2E1, Canada}
\author[0000-0002-9181-1161]{Annie~Hughes}
\affiliation{IRAP, Universit\'e de Toulouse, CNRS, CNES, UPS, (Toulouse), France} \author[0000-0001-6551-3091]{Kathryn Kreckel}
\affiliation{Astronomisches Rechen-Institut, Zentrum f\"{u}r Astronomie der Universit\"{a}t Heidelberg, M\"{o}nchhofstra\ss e 12-14, 69120 Heidelberg, Germany}
\author[0000-0002-6118-4048]{Sharon E. Meidt}
\affil{Sterrenkundig Observatorium, Ghent University, Krijgslaan 281-S9, 9000 Gent, Belgium}
\author[0000-0002-1370-6964]{Hsi-An Pan}
\affiliation{Department of Physics, Tamkang University, No.151, Yingzhuan Road, Tamsui District, New Taipei City 251301, Taiwan}
\author[0000-0002-0472-1011]{Miguel~Querejeta}
\affiliation{Observatorio Astron\'{o}mico Nacional (IGN), C/Alfonso XII, 3, E-28014 Madrid, Spain}
\author[0000-0002-3933-7677]{Eva Schinnerer}
\affiliation{Max-Planck-Institut f\"{u}r Astronomie, K\"{o}nigstuhl 17, D-69117, Heidelberg, Germany}
\author[0000-0003-2093-4452]{Cory M. Whitcomb}
\affiliation{Ritter Astrophysical Research Center, University of Toledo, Toledo, OH 43606, USA}



\begin{abstract}
Ratios of polycyclic aromatic hydrocarbon (PAH) vibrational bands are a promising tool for measuring the properties of the PAH population and their effect on star formation.  The photometric bands of the MIRI and NIRCam instruments on JWST provide the opportunity to measure PAH emission features across entire galaxy disks at unprecedented resolution and sensitivity. 
Here we present the first results of this analysis in a sample of three nearby galaxies: NGC~628, NGC~1365, and NGC~7496.  Based on the variations observed in the 3.3, 7.7, and 11.3~$\mu$m features, we infer changes to the average PAH size and ionization state across the different galaxy environments.  High values of \ratiose and low values of \ration are measured in \hii~regions in all three galaxies.
This suggests that these regions are populated by hotter PAHs, and/or that the PAH ionization fraction is larger.
We see additional evidence of heating and/or changes in PAH size in regions with higher molecular gas content as well as increased ionization in regions with higher \halpha intensity.
\end{abstract}



\section{Introduction}
\label{SecIntro}
The mid-infrared (mid-IR) emission features at 3.3, 6.2, 7.7, 8.6, 11.3, 12.6, and 17~$\mu$m first seen in Milky Way star--forming regions \citep[][]{Gillett1973, Merrill1975} and later observed in the spectrum of star-forming galaxies \citep[e.g.,][]{Smith2007} have been attributed to the vibrational modes of large carbonaceous molecules \citep[][]{Allamandola1989, DL2001,tielens2008}, referred to as polycyclic aromatic hydrocarbons \citep[PAHs,][]{LP1984, Allamandola1985}, or in some models, to the same modes of surface H atoms in the aromatic mantle of dust grains \citep[][]{THEMIS2017}. PAHs are important in a variety of processes in the ISM, including photoelectric heating of neutral gas \citep{bakes1994,wolfire1995,weingartner2001} and the formation of H$_2$ \citep{lepage2009,lebourlot2012}. 

The characteristics of these molecules, including their size, charge, hydrogenation, and structure, determine the efficiency of these processes \citep[e.g.\ the charge state of PAHs determines the efficiency of the photoelectric effect;][]{tielens2008}, and set the relative strength of the various mid-IR vibrational bands \citep{maragkoudakis2020}.
Intensive laboratory and theoretical work continues to identify the contributing species to each of these emission features or complexes of features \citep[e.g.][]{Boersma2014PAHdb, Bauschlicher2018PAHdb, Mattioda2020, kerkeni2022}. Observations of PAH band ratios can use this information to infer properties like average size and ionization by modeling the distribution of PAHs that contribute to the observed emission \citep[e.g.][]{Draine2021,rigopoulou2021,maragkoudakis2022,kerkeni2022}. 

Some of these features are now generally agreed to be produced by specific types of PAHs. For instance, the 3.3~$\mu$m feature is characteristic of small, neutral PAHs, while the 7.7~$\mu$m complex traces vibrational bands produced by larger, positively charged ions, and the 11.3~$\mu$m is representative of a population of grains that are both larger and neutral \citep[e.g.][]{Galliano2008, Boersma2016, Boersma2018, maragkoudakis2020, Draine2021, rigopoulou2021, maragkoudakis2022}.
The ratios of these bands can therefore constrain quantities such as the relative amounts of small to large PAHs (in terms of number of carbon atoms) and the fraction of ionized to neutral PAHs.

Determining how these ratios vary across a galaxy can also provide important context for how the PAHs are influencing the surrounding ISM.  For example, higher fractions of ionized PAHs could lead to a decrease in the photoelectric heating efficiency, as highly charged grains will have higher ionization potentials \citep{Tielens2008Review}.  This has been observed in variations in the relative strength of far-infrared cooling lines compared to total infrared luminosity as a function of PAH feature ratios, suggesting the properties of the PAHs influence the balance of ISM heating and cooling \citep{Croxall2012, McKinney2021, Sutter2022}. As both a source and sink for free electrons, PAHs also play a large role in setting the ionization balance \citep{Li2020Review, bakes1994}. Understanding the properties of the PAHs as traced by band ratios will therefore help establish how these important molecules influence ISM conditions.

The \textit{James Webb Space Telescope} (JWST) opens a new era in PAH studies by simultaneously offering a diverse suite of filters covering many key PAH features, and offering unprecedented resolution in the mid-IR. 

While previously restricted to Local Group galaxies \citep[see e.g.][]{Sandstrom2012}, it is now possible to directly probe these ratios across the disks of nearby galaxies, at small physical scales (tens of parsecs), with JWST.
With the NIRCam instrument \citep[][]{rieke2005} covering the 3.3~$\mu$m feature, and the MIRI instrument \citep[][]{rieke2015} covering the 7.7 and 11.3~$\mu$m complexes, we can investigate the properties of PAHs across the extent of $D\lesssim20$~Mpc galaxies at $\sim10-50$ pc resolution.  The sensitivity of JWST further enables measurements of the PAH properties in multiple phases of the ISM (e.g.\ diffuse atomic gas, molecular clouds without massive star forming regions, molecular clouds with associated \hii\ regions, diffuse ionized gas), not just the bright star--forming complexes that were visible to Spitzer IRS.

In this letter we deploy this new capability to measure the resolved variations in PAH band ratios in three of the first targets of the PHANGS-JWST survey \citep{LEE_PHANGSJWST}. We make maps of the \ratioss, \ratiose and \ration over wide areas to trace how the properties of the PAHs vary in comparison with multiwavelength tracers of the phase of the ISM.

\section{Data}
\label{SecData}
\subsection{JWST-MIRI maps tracing the $7.7$ and $11.3~\mu$m bands}
\begin{figure*}
    \centering
    \includegraphics[width=\textwidth]{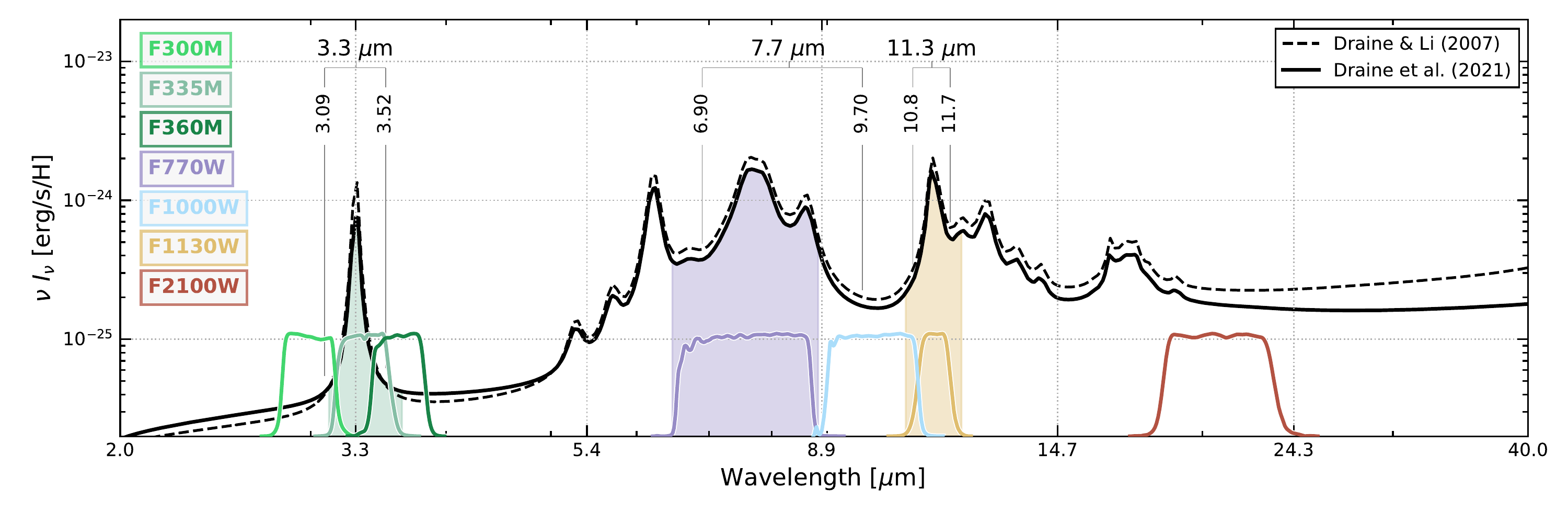}
    \caption{Visual representation of some of the NIRCam, and the MIRI filters, with their normalized transmission curves, in color, and associated integrated area for the three main filters used in this study.
    The vertical lines and the shaded regions mark the wavelength cuts for the clip points defined in \citet[][]{Draine2021} that they used to integrate the feature luminosities. The associated complexes are labels on top.
    The black solid and dashed lines are representative mid-IR spectra from \citet[][$U=1.0$, $b_{\rm C} = 0.05$]{DL07}, and \citet[][`mMMP' with $U=1.0$, `st' ionization and `std' size]{Draine2021}, for comparison.}
    \label{FigFilters}
\end{figure*}

We use the MIRI F770W and F1130W maps of three early targets in the PHANGS-JWST Treasury program \#2107 (PI: Lee): NGC~628, NGC~1365, and NGC~7496. Description of the data reduction for the MIRI images can be found in \citet{LEE_PHANGSJWST}. Since we are interested in the diffuse PAH emission throughout the maps, the background level and the uncertainty in the maps are critical.  The background level was tied to larger archival {\em Spitzer} and WISE maps of the targets with well-covered ``off'' galaxy regions \citep[see Appendix A and B in][]{LEROY1_PHANGSJWST}.
As shown in Figure~\ref{FigFilters}, the F770W and F1130W were ideally designed to sample the 7.7 and 11.3~$\mu$m PAH emission complexes. The filter transmission curves are shown as colored lines with labels on the right, and the shaded areas help visualize the fraction of each complex that is included in the filters used here.
For visual comparison purposes, we indicate with vertical lines the `clip points' used in \citet[][]{Draine2021}.
In this Letter, the F770W and F1130W values we use are instrument's filter-integrated fluxes.

All maps were convolved to the point spread function (PSF) of the F1130W filter (${\rm FWHM} \sim 0.36''$) using a standard convolution procedure, with kernels generated following the approach of \citet[][]{Aniano2011}. All maps were also aligned to the same astrometric grid, with pixels size $\sim 0.11''$.

Although the F770W and F1130W bands are dominated by the 7.7~$\mu$m and 11.3~$\mu$m PAH emission features, both can also include some hot dust and/or continuum emission, especially towards bright \hii~regions. 
\citet{Whitcomb2022} used PAHFIT \citep[][]{Smith2007} to show that the \textit{Spizter}/IRS spectra of the star-forming regions in the SINGS sample, which are among the brightest in these galaxies, may have $\sim 30\%$ continuum contamination in F1130W and $\sim 10\%$ in F770W, although these measurements rely on models of mid-IR emission that will likely be updated with new JWST results. 
Note that \hii~regions are the place where contamination from hot dust is the highest.
Naturally, a more detailed analysis and correction will be the subject of further study.

In addition to continuum contamination, it is also possible that silicate absorption at 9.7~$\mu$m could reduce the F1130W flux. 
\citet{Smith2007} show that this absorption should not play a large role in most star-forming galaxies.
Additionally, Groves et al. (submitted) also find that a median $E(B-V) \sim 0.2-0.3$ for \hii~regions in the four targets in this work, a relatively low value.
At this point, we therefore do not considered silicate absorption in this early work.

\subsection{JWST-NIRCam maps tracing the 3.3~$\mu$m band}
We use the 3.3~$\mu$m emission maps from \citet[][]{SANDSTROM2_PHANGSJWST}.
They use a combination of the NIRCam F335M (measuring the main PAH emission feature), and the F300M plus F360M filters to remove starlight continuum. They base their continuum removal approach on the prescription from \citet[][]{Lai2020}, with additional corrections to avoid over-subtraction of the PAH emission in regions where the map is ISM emission dominated rather than stellar continuum dominated in the F360M filter\footnote{Note that an additional factor would be required to convert F335M$_{\rm PAH}$ from MJy~sr$^{-1}$ into an integrated band intensity \citep[c.f.][]{Lai2020}.}. \citet{Lai2020} show contamination from emission lines is negligible in the F335M band. For full details on the removal of contamination from starlight in the F335M band, we refer to \citet{SANDSTROM2_PHANGSJWST}, which includes robust discussion of any potential remaining contamination which could artificially raise the F335M$_{\rm PAH}$. We refer to these final maps that isolate the emission from the 3.3~$\mu$m emission complex as F335M$_{\rm PAH}$.
We convolve these maps to the F1130W PSF using the same approach as the other maps, as described above.

\subsection{Ancillary data}
We combine CO, \hone, and \halpha measurements to trace the conditions of the ISM in our targets. 
We use the same CO/\hone/\halpha data as the companion letter on PAH fraction \citep{CHASTENET1_PHANGSJWST}.
We trace the molecular gas content using CO data from the PHANGS-ALMA survey \citep[][]{PHANGSALMA2021Pipeline, PHANGSALMA2021Survey}. We convert ${\rm ^{12}CO~(2-1)}$ measurements to H$_2$ surface density using the Milky Way CO-to-H$_2$ conversion factor $\alpha_{\rm CO} = 4.35$~M$_\odot$~pc$^{-2}$~(K~km~s$^{-1})^{-1}$, and a $^{12}\mathrm{CO}(2-1)$ to $^{12}\mathrm{CO}(1-0)$ line ratio $R_{21} = 0.65$ \citep[][]{denBrok2021, Leroy2022}.
We use \hone data from the PHANGS-MeerKAT survey (C. Eibensteiner et al., in prep) and THINGS \citep[][]{Walter2008} surveys, assuming optically thin 21-cm emission, and using a conversion factor of $0.020$~M$_\odot$~pc$^{-2}$~(K~km~s$^{-1})^{-1}$ \citep[][see also \citeauthor{Walter2008} \citeyear{Walter2008}]{Leroy2012}. We do not convolve the JWST and other datasets to match the $11''-15''$ \hone resolution. We assume that the \hone is smoothly distributed below the MeerKAT and THINGS resolution, as suggested by observations of more nearby targets \citep{leroy2013}.
There are currently no available 21-cm observations for NGC~1365, so we assume it to have a uniform neutral gas surface density of 8~M$_\odot$~pc$^{-2}$ everywhere, based on the observed flatness of the atomic gas distribution in the disks of galaxies \citep[e.g.][]{Schruba2011, Bigiel2012, KennicuttReview, Wong2013}.
We trace ionized gas using the \halpha data (not extinction corrected) from the PHANGS-MUSE observations \citep[][]{PHANGSMUSE2022Survey}.

\subsection{Noise properties and masks}
\label{SecNoise}
After convolution to the F1130W PSF, we measure a background standard deviation of $\sim 0.065$~\mjysr in the MIRI F770W map of NGC7496. It is the only target that provides sufficiently empty space to measure a background without significant contamination from the source.
We use that value to perform signal-to-noise (S/N) cuts in all three galaxies, and we exclude pixels with ${\rm S/N \leq 3}$ in the MIRI maps.
All MIRI maps have similar noise from the pipeline, verified with moderate masking, so it therefore is reasonable to assume the convolved maps shows similar noise. We consequently base our noise estimate off the convolved map of NGC~7496.

After convolution, we measure a noise in the F335M$_{\rm PAH}$ maps of $\sim 0.014$~\mjysr\ in NGC~7496. Similarly, we use that value in all three targets to perform cuts and remove pixels with ${\rm S/N \leq 3}$ in the F335M$_{\rm PAH}$ maps.
Figures~\ref{FigRatioNGC628} through \ref{FigRatioNGC7496} show the remaining pixels in the three targets.

Since NGC~1365 and NGC~7496 have saturated central sources that cause contamination of the surrounding maps by the PSF wings, we use the instrument PSF of the F1130W filter to mask the saturated and contaminated pixels in these galaxy centers.

\section{Band ratios}
\label{sec:bandratio}
\begin{figure*}
    \centering
    \includegraphics[width=\textwidth, clip, trim={3cm 1cm 3cm 0}]{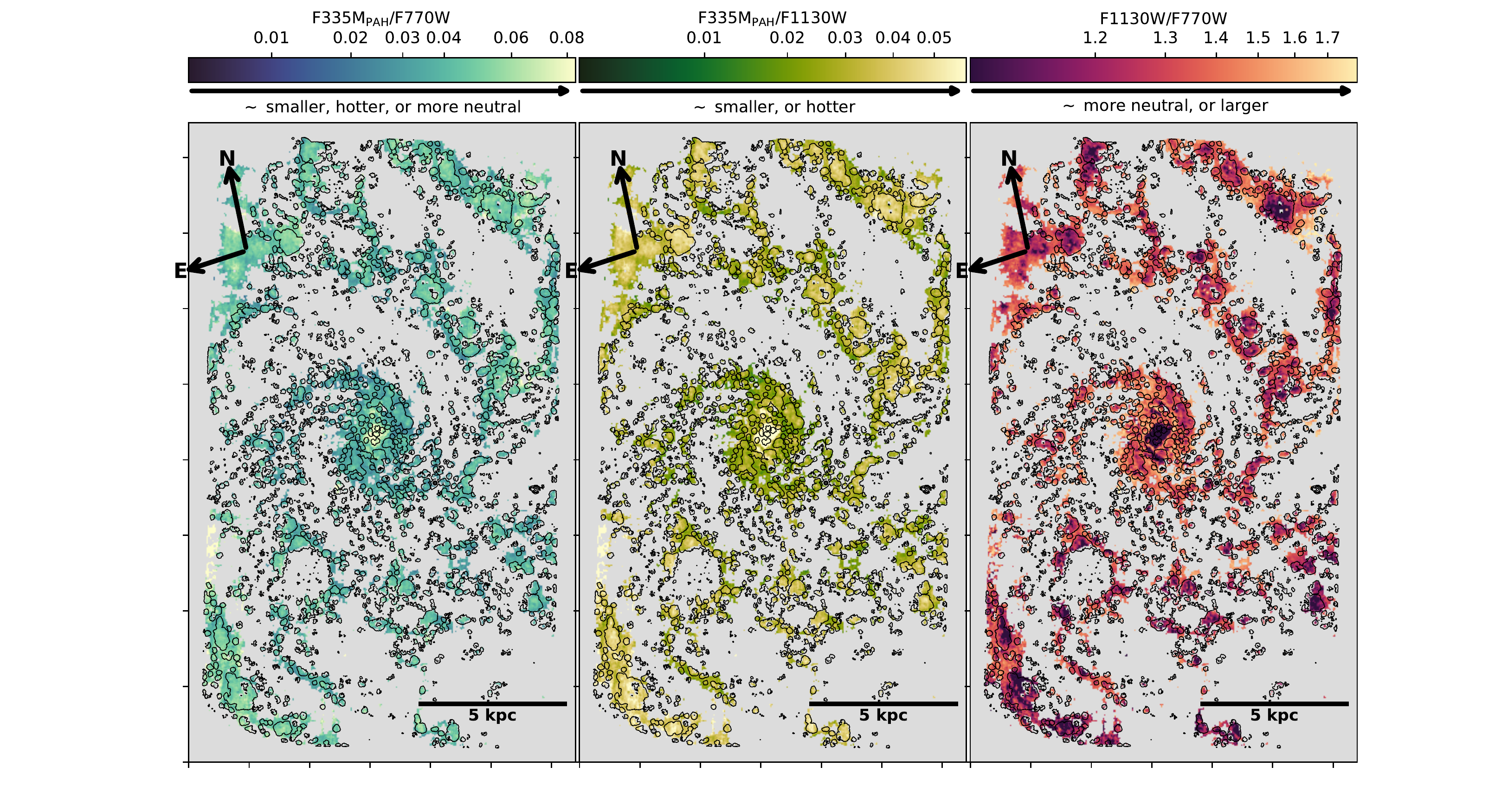}
    \caption{PAH band ratios in NGC~628. The maps show all pixels passing ${\rm S/N \geq 3}$ in all bands (see Section~\ref{SecNoise}).
    The F335M$_{\rm PAH}$ refers to the 3.3~$\mu$m maps developed by \citet[][]{SANDSTROM2_PHANGSJWST}.
    The contour show the \hii~regions from the nebula catalog by \citet[][]{GROVES_HIICAT, PHANGSMUSE2022HII}.
    (The maps were arbitrarily rotated from north-up to allow for better visualization.)
    }
    \label{FigRatioNGC628}
\end{figure*}

\begin{figure*}
    \centering
    \includegraphics[width=\textwidth, clip, trim={3cm 1.5cm 3cm 1cm}]{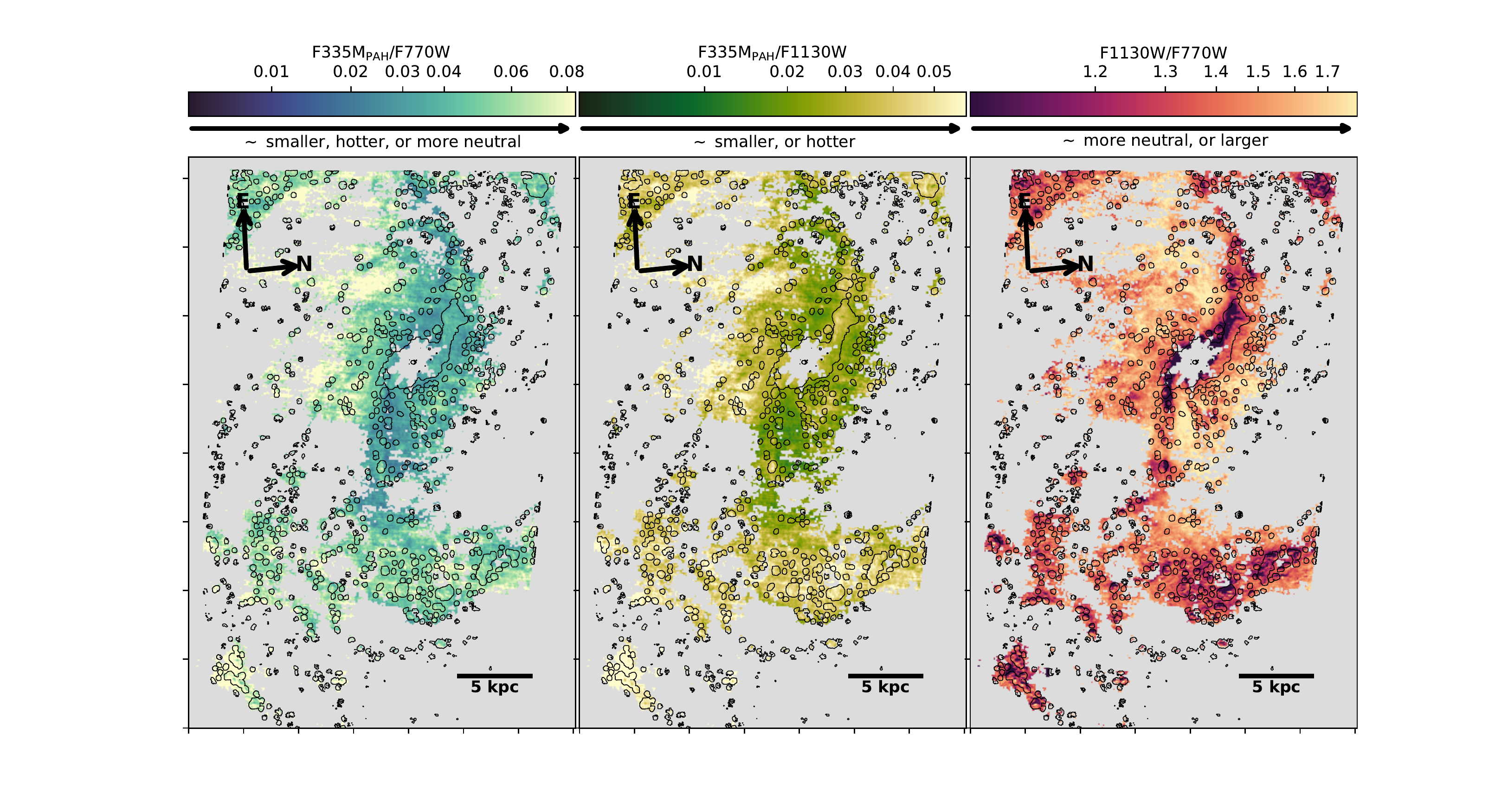}
    \caption{Same as Figure~\ref{FigRatioNGC628} for NGC~1365. The central pixels are masked due to saturation, using the PSF at F1130W. The visible stripes are due to $1/f$ noise in the NIRCam bands. 
    (The maps were arbitrarily rotated from north-up to allow for better visualization.)}
    \label{FigRatioNGC1365}
\end{figure*}

\begin{figure*}
    \centering
    \includegraphics[width=\textwidth, clip, trim={3cm 2cm 3cm 1.25cm}]{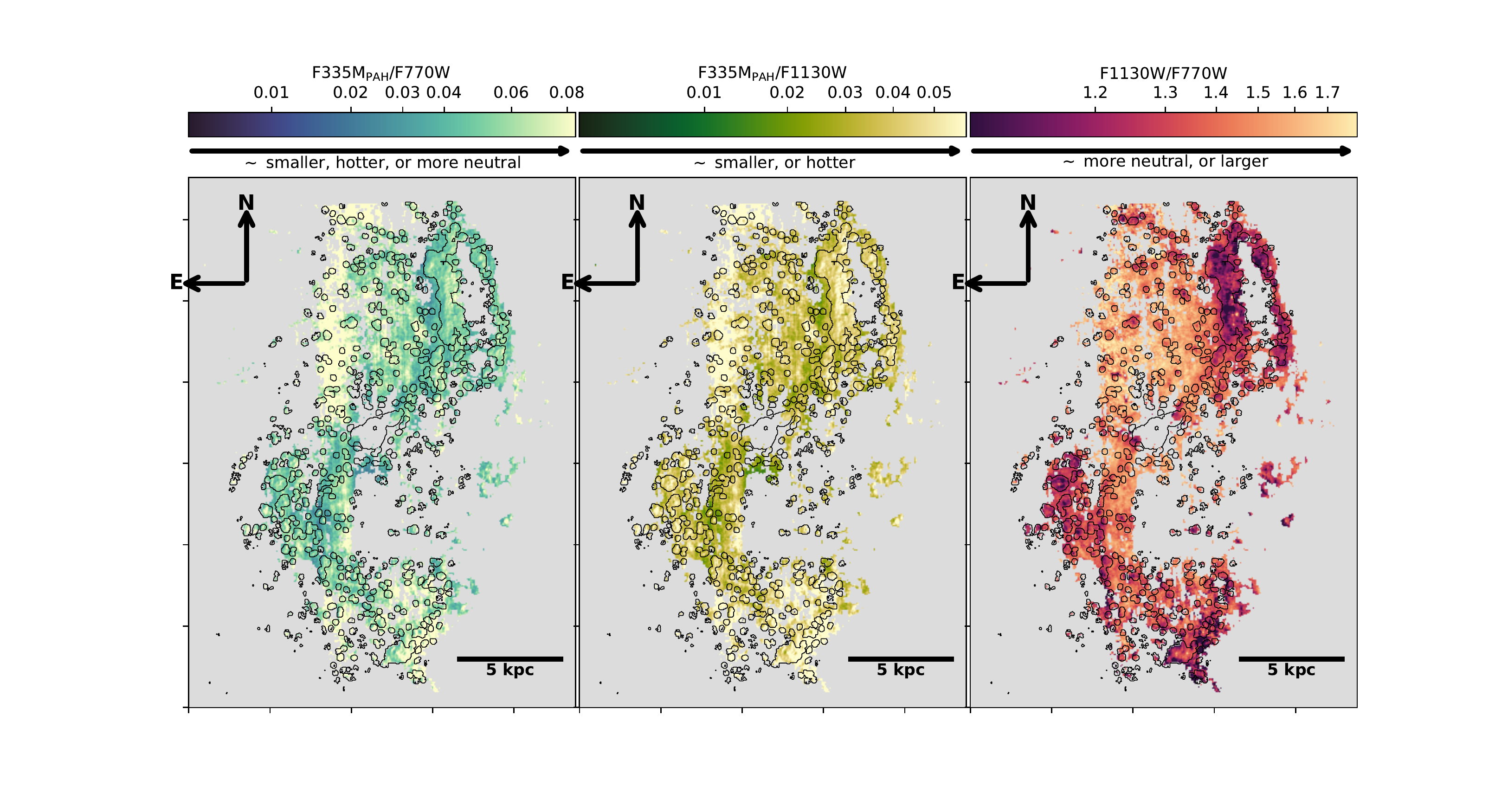}
    \caption{Same as Figure~\ref{FigRatioNGC628} for NGC~7496. The central pixels are masked due to saturation, using the PSF at F1130W. The visible stripes are due to $1/f$ noise in the NIRCam bands.}
    \label{FigRatioNGC7496}
\end{figure*}

We present maps of the \ratioss, \ratiose and \ration band ratios in NGC~628 (Figure~\ref{FigRatioNGC628}), NGC~1365 (Figure~\ref{FigRatioNGC1365}), and NGC~7496 (Figure~\ref{FigRatioNGC7496}). The color scale for each band ratio is the same across the three figures, with limits chosen using the 5$^{\rm th}$ and 95$^{\rm th}$ percentiles of all values.
We overlay the \hii~regions cataloged in \citet[][]{GROVES_HIICAT, PHANGSMUSE2022HII} with black contours.


Overall, each ratio shows variations both within and across galaxies, spanning values within a factor of 4--5 for \ratioss and \ratiose, and less than a factor of 2 in the \ration maps.
Table~\ref{TabData} lists the mean, median and ${\rm 16^{th}-84^{th}}$ percentile ranges of the three ratios in each galaxy. These values are also well-illustrated in Figure~\ref{FigPanel}.
The mean and median values in \ratioss and \ratiose increase from NGC~628, to NGC~1365, and to NGC~7496 which shows the highest ratios. The ${\rm 16^{th}-84^{th}}$ percentile ranges between NGC~628 and NGC~7496 barely overlap, indicating a real offset between these two galaxies in the \ratioss and \ratiose ratios.
On the other hand, all three galaxies show relatively similar mean and median values of the \ration ratio, with significant overlap in ${\rm 16^{th}-84^{th}}$ ranges.
This suggests that, in terms of band ratios, the main difference between each galaxy is driven by the emission from the 3.3~$\mu$m complex, traced by F335M$_{\rm PAH}$ \citep[see also][]{SANDSTROM2_PHANGSJWST}.

\begin{table*}[]
    \centering
    \begin{tabular}{|c|c|c|c|c|c|c|c|c|c|}
        \hline
        \multirow{2}{*}{\textbf{Galaxy}} & \multicolumn{3}{c|}{\textbf{F335M$_{\rm PAH}$/F770W}} & \multicolumn{3}{c|}{\textbf{F335M$_{\rm PAH}$/F1130W}} & \multicolumn{3}{c|}{\textbf{F1130W/F770W}} \\
        & Mean & Median & ${\rm 16^{th}-84^{th}}$ perc. & Mean & Median & ${\rm 16^{th}-84^{th}}$ perc. & Mean & Median & ${\rm 16^{th}-84^{th}}$ perc. \\
        \hline
        NGC~628 & 0.043 & 0.039 & $0.028 - 0.054$ & 0.032 & 0.029 & $0.02 - 0.042$ & 1.38 & 1.37 & $1.20 - 1.56$ \\
        NGC~1365 & 0.061 & 0.053 & $0.035 - 0.084$ & 0.042 & 0.037 & $0.023 - 0.058$ & 1.48 & 1.48 & $1.27 - 1.69$ \\
        NGC~7496 & 0.067 & 0.057 & $0.040 - 0.092$ & 0.049 & 0.041 & $0.029 - 0.065$ & 1.41 & 1.40 & $1.22 - 1.60$ \\
        \hline
    \end{tabular}
    \caption{Means, medians and ${\rm 16^{th}}$ and ${\rm 84^{th}}$ percentiles for the three ratios \ratioss (left), \ratiose (center), and \ration (right), in each galaxy. While the ${\rm 16^{th}-84^{th}}$ percentile ranges of \ration overlap between targets, the other ratios with F335M$_{\rm PAH}$ seem to show more distinct values.}
    \label{TabData}
\end{table*}

The ratio maps show higher values of the \ratioss and \ratiose ratios in \hii~regions.
This is clearly visible in NGC~628 and NGC~7496, and slightly less obvious in NGC~1365 though still distinguishable. 
This indicates enhanced 3.3~$\mu$m emission in these regions compared to the other two PAH features. 
In all galaxies, the \ration maps show darker colors in \hii~regions, pointing to increased average PAH ionization likely driven by the proximity of young stellar populations.  This is similar to what is seen in \citet[][]{DALE_PHANGSJWST}, which examines the properties of the PAHs within compact stellar clusters and finds a positive correlation between cluster age and PAH ionization.
In the galaxy NGC~1365, darker colors in the right panel of Figure~\ref{FigRatioNGC1365} (i.e., lower values of \ration), tracing more ionized PAHs, can be found in the dust lanes. This is also visible in NGC~7496.

The relative prominence of the 3.3~$\mu$m feature in \hii~regions could suggest that PAHs in these regions are smaller, and/or hotter than in the surrounding diffuse ISM.  
The emission in the 3.3~$\mu$m is usually attributed to small PAHs. In this case, the higher \ratioss and \ratiose would suggest smaller PAHs in \hii regions. 
But the seemingly higher fraction of smaller PAHs in harsh environment is not supported by theoretical studies, which have shown that smaller PAHs have shorter lifetimes in a hot gas \cite[e.g.,][]{Micelotta2010}.
However, it is important to note that in the vicinity of \hii~regions with harder radiation fields, PAHs will be hotter due to an increase in the average energy of the photons. 
An increase in dust temperature naturally shifts the power in the emission spectrum towards shorter wavelengths; in this case, hotter PAHs would lead to more power in the 3.3~$\mu$m feature, interpreted as an \emph{apparent} shift towards smaller sizes \citep{Draine2021}.
This shift is clearly demonstrated in Figure 21, panel d of \citet[][]{Draine2021}, in which the value of F$_{\rm clip}(3.3)$/F$_{\rm clip}(11.2)$ irradiated by a Galactic radiation field (mMMP) is approximately a factor of two lower than the F$_{\rm clip}(3.3)$/F$_{\rm clip}(11.2)$ irradiated by a young stellar population (BC03, 3Myr).  We note that the shift reported by the models is smaller than the shift we see in the \hii regions across the galaxies presented here, suggesting potential additional factors not accounted for in the \citet{Draine2021} models.

The combined analysis of these three galaxies also indicates that the \ration ratios are lowest in \hii~regions, the star-forming arms, and gas and dust lanes that may include shocked gas. 
This argues that the fraction of ionized PAHs surrounding sites of active star formation are is on average higher than in the diffuse ISM.
This is further discussed in \citet[][]{EGOROV_PHANGSJWST}, which examines the PAH emission in \hii~regions.  The relative consistency of the ratios in the interarm ISM where detected, despite sampling three different galaxies, implies that PAHs permeate the diffuse ISM and have similar properties throughout this ISM phase.

Note that we attribute the high \ratioss and \ratiose in the inner 200~pc ($\sim 4''$) of NGC~628 to the very low ISM contribution in that region, and not a change in PAH properties
\citep[e.g.,][]{Dale2006, HOYER_PHANGSJWST}.

\section{Variation of band ratios with ISM environment}
\label{sec:ISM}
\begin{figure*}
    \centering
    \includegraphics[width=\textwidth]{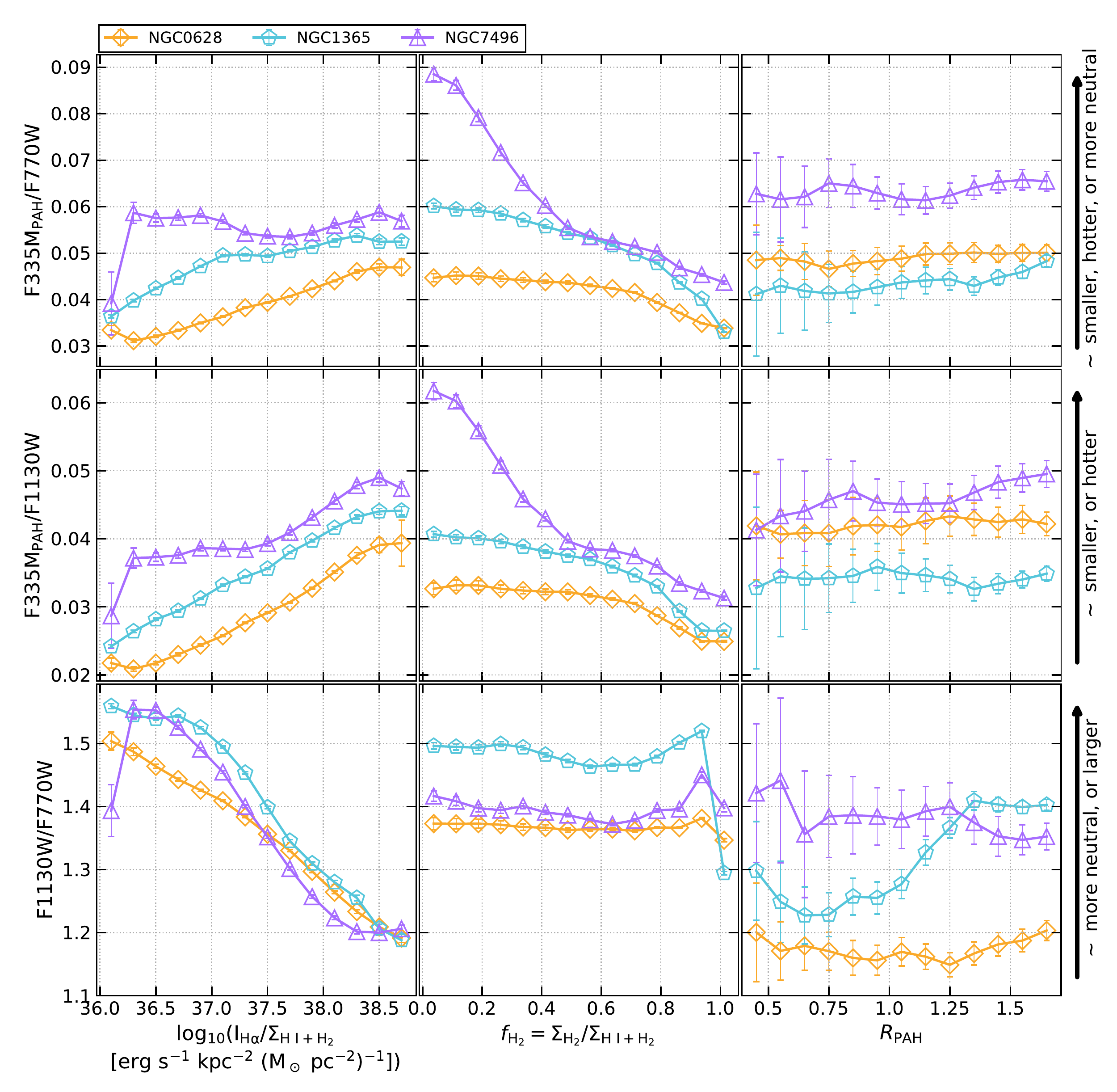}
    \caption{Running medians of the \ratioss (top), \ratioss (middle), and \ration (bottom), in bins of \hagasratio (left), fraction of molecular gas, $f_{\rm H_2}$ (right), and the abundance ratio \rpah~=~(F700W+F1130W)/F2100W.}
    \label{FigPanel}
\end{figure*}

In Figure~\ref{FigPanel}, we show the variations of the three ratios, \ratioss (top row), \ratiose (middle row), and \ration (bottom row), as a function of \hagasratio in units of ${\rm erg~s^{-1}~kpc^{-2}~(M_\odot~pc^{-2}})^{-1}$ (left column), and fraction of molecular gas (right column). 
The symbols are the medians in each bin of the x-axis, and the error bars show 3 standard errors of the mean.

All ratios show clear variations with the ionized gas content, as shown in the first column.
The first two ratios are sensitive to the global size of the PAHs, and increase with \hagasratio. This suggests that the PAH population decreases to lower sizes as the ionized gas becomes more prominent, though heating could also drive this observed change (as discussed in Section \ref{sec:bandratio}). 
The \ration ratio, on the other hand, decreases at high values of \hagasratio, which suggests that the PAHs are more ionized with a higher fraction of ionized gas to total gas. 
Both these trends are consistent with the expected first order effect from hot gas, that should be able to fragment PAHs or kick electrons. However, there is also a known negative effect of ionized gas on the global PAH abundance \citep[e.g.][]{Micelotta2010, Chastenet2019}.
The last column of Figure~\ref{FigPanel} shows the variations of the ``abundance ratio'', $R_{\rm PAH} = {\rm (F770W+F1130W)/F2100W}$ tracing the global PAH abundance\footnote{To first order, (7.7+11.3)/21 traces PAH abundance, as was seen by the equivalent 8/24~$\mu$m that was frequently used in Spitzer observations \citep[e.g.,][]{Engelbracht2006, Engelbracht2008, DL07}.
In our early results, we are focused on observed ratios, rather than modeling.  Any exact relationship between the PAH fraction and the ratio used to measure $R_{\rm PAH}$ will be left to future studies.}
\citep[see also Letters from][in this Issue]{CHASTENET1_PHANGSJWST, EGOROV_PHANGSJWST}. In these panels, neither of the ratios shows a strong trend with \rpah. 
\citet[][]{CHASTENET1_PHANGSJWST} show that \rpah clearly decreases with \hagasratio, but no clear variation is seen with size or ionization tracers.

The second column shows the variations of the three ratios with the fraction of molecular gas, $f_{\rm H_2} \equiv \Sigma_{\mathrm{H_2}} / (\Sigma_{\rm H{\sc I}}+\Sigma_{\mathrm{H_2}})$.
Both ratios tracing the size of PAHs (\ratioss, and \ratiose) show a steep decrease with $f_{\rm H_2}$ in NGC~7496, and a much shallower trend in NGC~0628 and NGC~1365.
In the bottom-center panel, however, while all curves show similar trends, only NGC~1365 differs from NGC~628, while NGC~7496 show identical values over a large range of $f_{\rm H_2}$. 
While NGC~1365 clearly has a higher CO content, NGC~628 and NGC~7496 show similar $L_{\rm CO}$ values \citep[][]{PHANGSALMA2021Survey}.
Additionally, note that both NGC~1365 and NGC~7496 host an AGN in their center.
The variations shown in the center column could indicate that: (1) the presence of an AGN has a stronger effect on the size of PAHs rather than their ionization, and (2) the CO content will have the opposite effect, with a higher impact on ionization rather than size.
However, these two Seyfert galaxies show different trends, making it difficult to clearly determine the effect of an AGN on the PAH population at this stage.

\begin{figure*}
    \centering
    \includegraphics[width=\textwidth, clip, trim={1cm 1cm 3cm 0.5cm}]{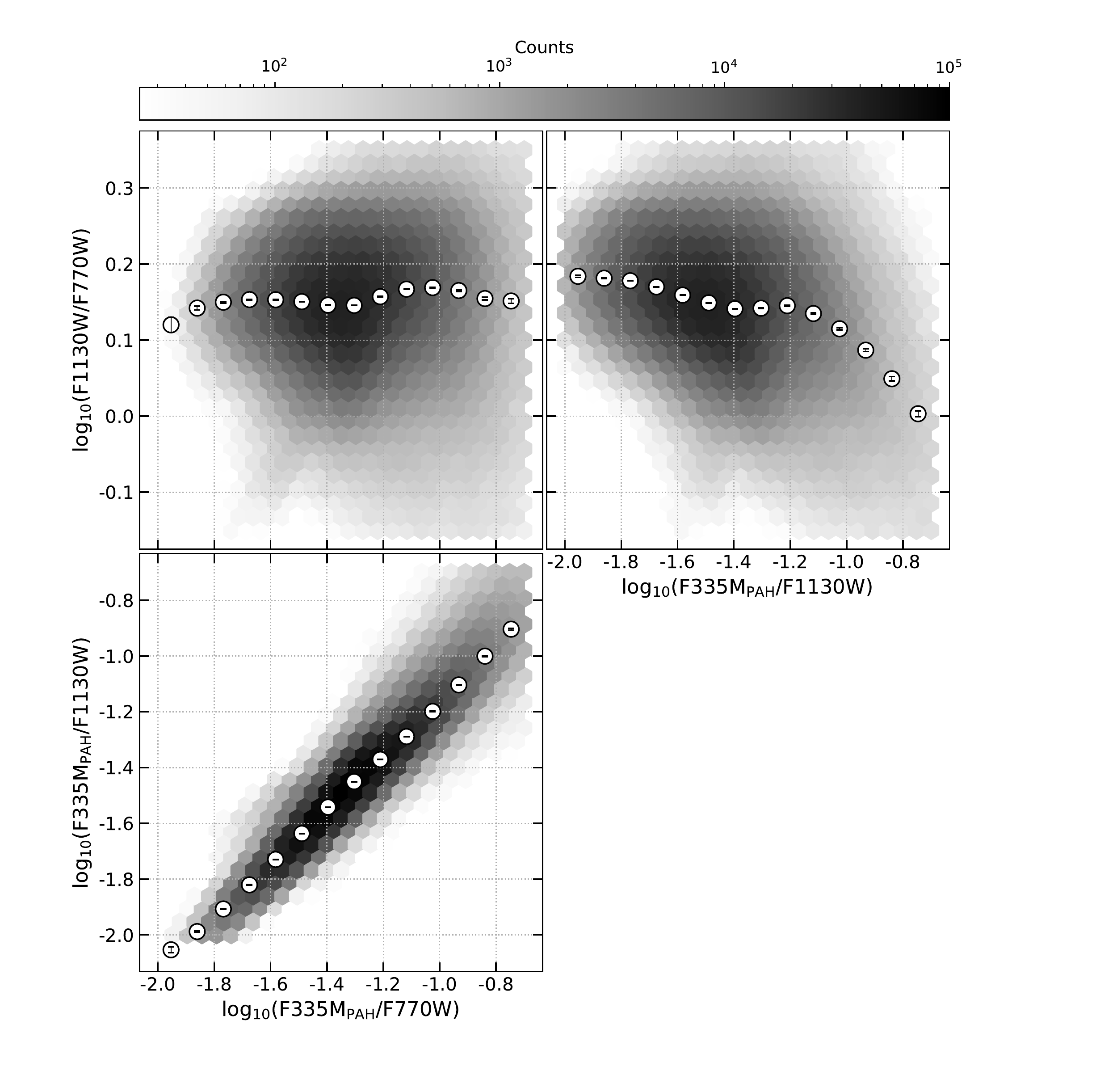}
    \caption{2D histograms of the \ratioss vs \ration (top), and \ratioss vs \ratiose (bottom). Both ratios in the bottom panel trace the size of PAHs, and show a clear correlation. 
    In the top panel, the x-axis traces size, while the y-axis traces ionization.}
    \label{Fig2DHists}
\end{figure*}

In Figure~\ref{Fig2DHists} we show the 2D histograms of all three ratios with each other, with the associated running medians and error bars showing 3 standard errors of the mean. 
In the bottom panel, the \ratioss and \ratiose appear clearly correlated, which is expected as they are both tracers of the global size distribution of PAHs\footnote{We note that the bulk of the \ratiose distribution is lower than the one found in \citet[][]{SANDSTROM2_PHANGSJWST}. This is likely due to the convolution to F1130W resolution, which leads to including more pixels in the diffuse regions, with lower F335M$_{\rm PAH}$ values.}.
In the top row, the \ration as a function of \ratioss (left) shows a rather flat distribution. In the right panel, \ration seems to gently decrease with \ratiose (though the large scatter can be consistent with a flat trend). 
The lack of a clear trend between \ration and the other ratios is interesting as it suggests that ionization and size do not strongly vary together.
A clear interpretation of the difference between the two panels in the top row requires a clear understanding of the different populations traced by \ratioss and \ratiose.

\section{Conclusions}
\label{sec:con}
Data from JWST provide exciting opportunities for understanding the ISM conditions of galaxies outside the Milky Way at $\sim$10 pc resolution. In particular, the MIRI and NIRCam instruments are able to capture the strength of important PAH features which can be used to assess the properties of these dust grains. Studies of the PAH population in nearby galaxies have shown the PAHs have an important impact on other evolution process, as they can be used as a star formation rate tracer and are heavily responsible for the photoelectric heating of the gas.
Through analysis of the ratio of the 11.3~$\mu$m to the 7.7~$\mu$m PAH features, changes in the ionization fraction of the PAHs can be determined, while the strength of the 3.3~$\mu$m compared to the 7.7~$\mu$m or 11.3~$\mu$m features provides information on the average PAH size.

We present maps of these ratios across nearby galaxies NGC~628, NGC~1365, and NGC~7496.  Based on the relative consistency of these ratio across each galaxy disk, we conclude that the population of PAHs is fairly uniform.  \hii~regions and star-forming spiral arms show lower 11.3/7.7 and higher \ratiose values, indicating the presence of smaller, highly charged grains in these regions.  The trends in these ratios are further discussed in \citet{DALE_PHANGSJWST}, where each ratio is measured in a sample of over 1000 stellar clusters.  In addition, the trends observed between the state of the gas measured by $f_{\rm H_2}$ and \hagasratio show that local gas conditions can greatly impact the sizes and ionization levels of the PAHs: more ionized (traced by \ration) and smaller PAHs (traced by \ratioss and \ratiose) are present in regions showing high \halpha intensity.

This work highlights the diagnostic power of JWST MIRI and NIRCam observations of nearby galaxies.  By probing previously inaccessible spatial scales and emission levels, we can better assess the physical mechanisms that link PAH grains and their local environment.  Further work to determine how these PAH feature ratios are influenced by stellar population age, gas--phase metallicity, and other galactic-scale and ISM properties will shed light on the delicate interplay between stars, gas, and dust in galaxies.


\section*{Acknowledgments}
We thank the anonymous referee for their careful reading and comments that helped improve the clarity of the paper.
This work was carried out as part of the PHANGS collaboration, associated with JWST program 2107. This work is based on observations made with the NASA/ESA/CSA JWST. 
Some/all of the data presented in this paper were obtained from the Mikulski Archive for Space Telescopes (MAST) at the Space Telescope Science Institute, which is operated by the Association of Universities for Research in Astronomy, Inc., under NASA contract NAS 5-03127.
The specific observations analyzed can be accessed via \dataset[10.17909/9bdf-jn24]{http://dx.doi.org/10.17909/9bdf-jn24}.
Based on observations collected at the European Southern Observatory under ESO programmes 094.C-0623 (PI: Kreckel), 095.C-0473,  098.C-0484 (PI: Blanc), 1100.B-0651 (PHANGS-MUSE; PI: Schinnerer), as well as 094.B-0321 (MAGNUM; PI: Marconi), 099.B-0242, 0100.B-0116, 098.B-0551 (MAD; PI: Carollo) and 097.B-0640 (TIMER; PI: Gadotti). 
This paper makes use of the following ALMA data: \linebreak
ADS/JAO.ALMA\#2012.1.00650.S, \linebreak 
ADS/JAO.ALMA\#2013.1.01161.S, \linebreak 
ADS/JAO.ALMA\#2015.1.00925.S, \linebreak 
ADS/JAO.ALMA\#2015.1.00956.S, \linebreak 
ADS/JAO.ALMA\#2017.1.00392.S, \linebreak 
ADS/JAO.ALMA\#2017.1.00766.S, \linebreak 
ADS/JAO.ALMA\#2017.1.00886.L, \linebreak 
ADS/JAO.ALMA\#2018.1.01651.S. \linebreak 
ADS/JAO.ALMA\#2018.A.00062.S. \linebreak 
ALMA is a partnership of ESO (representing its member states), NSF (USA) and NINS (Japan), together with NRC (Canada), MOST and ASIAA (Taiwan), and KASI (Republic of Korea), in cooperation with the Republic of Chile. The Joint ALMA Observatory is operated by ESO, AUI/NRAO and NAOJ.

JC acknowledges support from ERC starting grant \#851622 DustOrigin.
EJW acknowledges funding from the Deutsche Forschungsgemeinschaft (DFG, German Research Foundation) -- Project-ID 138713538 -- SFB 881 (``The Milky Way System'', subproject P1). 
HAP acknowledges support by the National Science and Technology Council of Taiwan under grant 110-2112-M-032-020-MY3.
MC gratefully acknowledges funding from the DFG through an Emmy Noether Research Group (grant number CH2137/1-1).
COOL Research DAO is a Decentralized Autonomous Organization supporting research in astrophysics aimed at uncovering our cosmic origins.
JMDK gratefully acknowledges funding from the European Research Council (ERC) under the European Union's Horizon 2020 research and innovation programme via the ERC Starting Grant MUSTANG (grant agreement number 714907).
TGW acknowledges funding from the European Research Council (ERC) under the European Union’s Horizon 2020 research and innovation programme (grant agreement No. 694343).
MB acknowledges support from FONDECYT regular grant 1211000 and by the ANID BASAL project FB210003.
KK, OE gratefully acknowledge funding from the Deutsche Forschungsgemeinschaft (DFG, German Research Foundation) in the form of an Emmy Noether Research Group (grant number KR4598/2-1, PI Kreckel).
FB would like to acknowledge funding from the European Research Council (ERC) under the European Union’s Horizon 2020 research and innovation programme (grant agreement No.726384/Empire).
MQ acknowledges support from the Spanish grant PID2019-106027GA-C44, funded by MCIN/AEI/10.13039/501100011033.
ER and HH acknowledge the support of the Natural Sciences and Engineering Research Council of Canada (NSERC), funding reference number RGPIN-2022-03499.
KG is supported by the Australian Research Council through the Discovery Early Career Researcher Award (DECRA) Fellowship DE220100766 funded by the Australian Government. 
KG is supported by the Australian Research Council Centre of Excellence for All Sky Astrophysics in 3 Dimensions (ASTRO~3D), through project number CE170100013. 
AKL gratefully acknowledges support by grants 1653300 and 2205628 from the National Science Foundation, by award JWST-GO-02107.009-A, and by a Humboldt Research Award from the Alexander von Humboldt Foundation.


\facilities{JWST (NIRCam, MIRI), MUSE, ALMA}

\bibliography{main,phangsjwst}{}
\bibliographystyle{aasjournal}



\end{document}